# Cosmic Modesty

Abraham Loeb

The richness of the universe teaches us modesty and guides us to search for both primitive and intelligent forms of life elsewhere without prejudice.

"There are many reasons to be modest", my mother used to say when I was a kid. But after three decades as an astronomer, I can add one more reason: the richness of the Universe around us.

Prior to the development of modern astronomy, humans tended to think that the physical world centers on us. The Sun and the stars were thought to revolve around the Earth. Although naive in retrospect, this is a natural starting point. When my daughters were infants, they tended to think that the world centers on them. Their development portrayed an accelerated miniature of human history. As they grew up, they matured and acquired a more balanced perspective.

Similarly, observing the sky makes us aware of the big picture and teaches us modesty. We now know that we are not at the center of the physical Universe, since the Earth orbits the Sun which circles around the center of the Milky Way galaxy, which itself drifts with a peculiar velocity of $\sim 0.001c$ relative the cosmic rest frame.

However, many people still believe that we might be at the center of the biological Universe, namely that life is rare or unique to Earth. In contrast, my working hypothesis, drawn from the above example of the physical Universe, is that we are not special in general, not only in terms of our physical coordinates but also as a form of life. Adopting this perspective implies that we are not alone. There should be life out there in both primitive and intelligent forms. This conclusion, implied by the principle of "cosmic modesty", has implications. If life is likely to exist elsewhere, we should search for it in all of its possible forms.

Our civilization reached an important milestone. We now have access to unprecedented technologies in our search for extraterrestrial life, be it primitive or intelligent. The search for primitive life is currently underway and well funded, but the search for intelligence is out of the mainstream of federal funding agencies. This should not be the case given that the only planet known to host life, the Earth, shows both primitive and intelligent life forms of it.

Our first radio signals leaked by now out to a distance of more than a hundred light years and we might soon hear back a response. Rather than being guided by Fermi's paradox: "where is everybody?" or by philosophical arguments about the rarity of intelligence, we should invest funds in building better observatories and searching for a wide variety of artificial signals on the sky. Civilizations at our technological level might produce mostly weak signals, and we will detect them only once we reach some exquisite sensitivity threshold. For example, a nuclear war on the nearest planet outside the solar system would not be visible even with our largest telescopes. But very advanced civilizations could potentially be detectable out to the edge of the observable Universe through their most powerful beacons[1]. The evidence for an alien civilization might not be in the traditional form of radio communication signals. Rather, it could involve detecting artifacts on planets through the spectral edge from solar cells[2], industrial pollution of atmospheres[3], artificial lights[4] or bursts of radiation from artificial beams sweeping across the sky for the propulsion of lightsails[5]. Finding the answer to the important question: "are we alone?" will change our perspective on our place in the Universe and will open up new interdisciplinary fields of research, such as astro-linguistics (how to communicate with aliens), astro-politics (how to negotiate with them for information), astro-sociology (how to interpret their collective behavior), astro-economics (how to trade space-based resources), and so on. We could shortcut our own progress by learning from civilizations that benefitted from a head start of billions of years.

There is no doubt that noticing the big picture taught my young daughters modesty. Similarly, the Kepler satellite survey of nearby stars allowed astronomers to infer that there are probably more habitable Earth-mass planets in the observable volume of the Universe than there are grains of sand on all beaches on Earth. Emperors or kings who boasted after conquering a piece of land on Earth resemble an ant that hugs with great pride a single grain of sand on the landscape of a huge beach.

Just over the past year, astronomers discovered a habitable planet, Proxima b, around the nearest star, Proxima Centauri[6], as well as three habitable planets out of seven around another nearby star[7] TRAPPIST-1 (and if life formed on one of the three, it was likely transferred to the others[8]). These dwarf stars, whose masses are 12% and 8% of the mass of the Sun respectively, will live for up to ten trillion years, about a thousand times longer than the Sun. Hence, they provide excellent prospects for life in the distant future, long after the Sun will die and turn into a cold white dwarf. I therefore advise my wealthy friends to buy real estate on Proxima b, as its value will likely go up dramatically in the future. But this also raises an important scientific question: "is life most likely to emerge at the present cosmic time near a star like the Sun?" By surveying the habitability of the Universe throughout cosmic history from the birth of the first stars 30 million years after the Big Bang to the death of the last stars in ten trillion years, one reaches the conclusion[9] that unless habitability around low mass stars is suppressed, life is most likely to exist near dwarf stars like Proxima Centauri or TRAPPIST-1 trillions of years from now.

The chemistry of "life as we know it" requires liquid water, but being at the right distance from the host star for achieving a comfortable temperature on the planet's surface is not a sufficient condition for life. The planet also needs to have an atmosphere. In the absence of an external atmospheric pressure, warming by starlight would transform water ice directly into gas rather than a liquid phase. The warning sign can be found next door: Mars has a tenth of the Earth's mass and lost its atmosphere. Does Proxima b have an atmosphere? If so, the atmosphere and any surface ocean it sustains, will moderate the temperature contrast between its permanent day and night sides. The *James Webb Space Telescope*, scheduled for launch in October 2018, will be able to distinguish between the temperature contrast expected if Proxima b is bare rock compared to the case where its climate is moderated by an atmosphere or an ocean[10]. Radio observations can inform us about the strength of Proxima b's magnetic field[11], which is also important for retaining its atmosphere.

A cosmic perspective about our origins would also contribute to a balanced worldview. The heavy elements that assembled to make the Earth were produced in the heart of a nearby massive star that exploded. A speck of this material takes form as our body during our life but then goes back to Earth (with one exception, namely the ashes of Clyde Tombaugh, the discoverer of Pluto, which were put on the New Horizons spacecraft and are making their way back to space). What are we then, if not just a transient shape that a speck of material takes for a brief moment in cosmic history on the surface of one planet out of so many? Despite all of this, life is still the most precious phenomenon we treasure on Earth. It would be amazing if we find evidence for "life as we know it" on the surface of another planet, and even more remarkable if our telescopes will trace evidence for an advanced technology on an alien spacecraft roaming through interstellar space.

*Abraham (Avi) Loeb is the Frank B. Baird Jr. Professor of Science at Harvard University. He serves as chair of the Harvard Astronomy Department, founding director of the Black Hole Initiative and director of the Institute for Theory & Computation at the Harvard-Smithsonian Center for Astrophysics, 60 Garden Street, Cambridge, Massachusetts 02138, USA.*
*e-mail: [aloeb@cfa.harvard.edu](aloeb@cfa.harvard.edu)*